# High energy and narrow linewidth $N_2$-filled hollow-core fiber laser at 1.4 μm

Lujun Hong,[1] Cuiling Zhang,[1] Joseph Wahlen,[2] Jose Enrique Antonio-Lopez,[2] Rodrigo Amezcua-Correa,[2] Christos Markos,[1,3,4] and Yazhou Wang[1,*]

[1]*DTU Electro, Technical University of Denmark, 2800 Kgs. Lyngby, Denmark*
[2]*CREOL, The College of Optics and Photonics, University of Central Florida, Orlando, Florida 32816, USA*
[3]*NORBLIS ApS, Virumgade 35D, 2830 Virum, Denmark*
[4]chmar@dtu.dk
[*]yazwang@dtu.dk

**In this work, we develop a high energy laser at 1.4 μm wavelength through the 1st order vibrational Raman Stokes generation in a nitrogen ($N_2$) filled nodeless anti-resonant hollow core fiber (ARHCF), pumped using a custom-made 1060 nm Yb-doped narrow-linewidth fiber laser with 3.7 ns pulse duration and 98 μJ pulse energy. The maximum Raman energy of 26.5 μJ is achieved at 15 bar $N_2$ pressure, corresponding to a quantum efficiency of 45% and peak power of 8.03 kW with 3.3 ns pulse width. This Raman laser has a narrow linewidth of 0.16 nm, and it is tunable over ~2 nm range by thermally tuning the laser diode seed wavelength of the pump fiber laser. This work provides a promising alternative for developing high energy and narrow-linewidth lasers outside the gain spectral regions of rare earth ion doped fibers.**

High energy and narrow linewidth lasers at 1.4 μm wavelength have important applications in gas spectroscopy, bio-imaging, material processing, and nonlinear wavelength conversion [1,2]. Conventional laser technologies at this wavelength include bismuth doped fiber lasers, Raman lasers, semiconductor lasers, and optical parametric oscillators (OPO). However, most of these laser technologies are limited to the (quasi-)continuous wave regime [3–8]. Boosting the pulse energy, achieving a narrow linewidth, and minimizing the footprint remain long-standing technological challenges for laser development in this spectral region. Consequently, high energy and narrow linewidth lasers at ~1.4 μm have been rarely reported [9–11], where pulse energy higher than microjoule levels have only been achieved by using OPO.

Raman Stokes frequency shift based on the recent advent of gas-filled anti-resonant hollow core fiber technology opens a promising way for wavelength conversion [12–17]. Compared to solid-core fibers, gases offer a significantly narrower Raman gain width and longer Stokes shift coefficient. The former fulfills the prerequisite of the generation of narrow laser linewidth [18,19], while the latter can convert the pump wavelength towards a broad spectral region such as ultraviolet and mid-infrared [14,17]. Combining gases with the ARHCF technology also offers the advantages of weak dispersion walk-off effect and high damage threshold, since the laser beam mainly propagates in the gas-filled hollow core region of ARHCF [20,21].

Several high-energy gas-filled hollow core fiber Raman lasers have been reported so far at different wavelengths spanning from UV to mid-infrared, where hydrogen, deuterium, methane, and carbon dioxide are the gas media widely used due to their high Raman gain and long Raman Stokes shift coefficient [14,18,22–26]. However, these gases have never been used for Raman laser generation at 1.4 μm because the required pump wavelength is outside the gain range of mature pump lasers (e.g. ytterbium(Yb)-doped laser @ ~1μm). On the other hand, the 1st order vibrational Raman Stokes of $N_2$ with a shift coefficient of 2331 cm$^{-1}$ can be used for wavelength conversion from ~1 μm to 1.4 μm. Nevertheless, there are very few results have been reported using $N_2$-filled hollow-core fiber [27,28]. And in these works, to enable the stimulated Raman scattering (SRS) process of $N_2$, they chose to use high energy ultrafast lasers as the pump, leading to a significant broadening of the Raman spectrum over tens nanometer due to the high pulse peak power.

In this work, we report a narrow linewidth and high-energy $N_2$-filled ARHCF Raman laser at 1.4 μm using a custom-made Yb-doped fiber laser as the pump. Figure 1(a) shows the schematic of the experimental setup. The pump fiber laser consists of a laser diode seed followed by two stages of polarization-maintained (PM) Yb-doped fiber pre-amplifiers and a PM Yb-doped fiber power amplifier. The seed laser is electrically modulated, to emit a linear polarized pulse train at a repetition rate of 1 kHz with 3.7 ns pulse width. After amplification, the pulse energy is boosted to 98 μJ [29], measured using an energy meter (PE9-ES-C, Ophir Optronics). The pump laser is coupled into the ARHCF through a plano-convex lens (LA1304-B, Thorlabs) with a coupling efficiency of ~80 %. A half-wave-plate (HWP) is placed before the input side of the ARHCF to adjust the polarization direction of the pump laser and therefore to optimize the vibrational Raman laser efficiency.

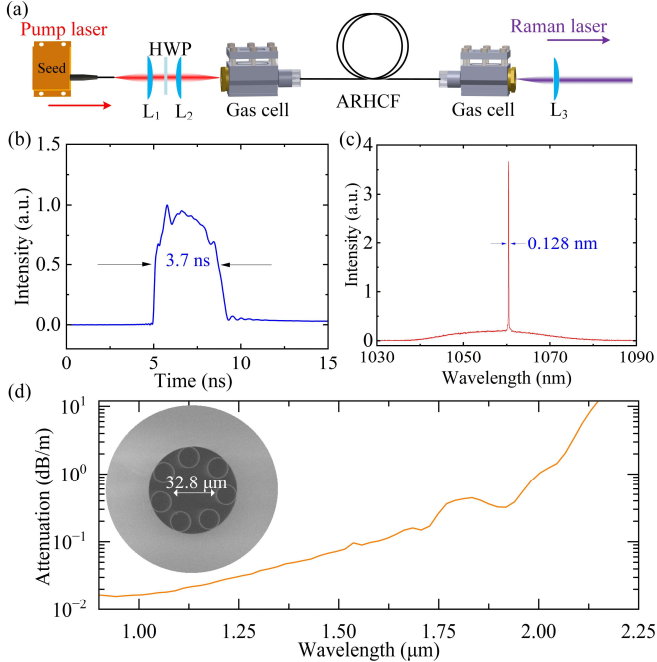

Fig. 1. (a) Experiment setup. (b) Pulse profile of the pump. (c) Spectrum of the pump. (d) The simulated attenuation spectrum of the ARHCF that is used in this work. The inset shows the cross-section of the ARHCF measured with a scanning electron microscope.

Figure 1(b) presents the pulse profile of the pump laser, measured by a 5 GHz near-IR photodetector (DET08C/M, Thorlabs) connected to an oscilloscope (6 GHz, Tektronix), showing a pulse duration of 3.7 ns. The spectrum of the pump laser is shown in Fig. 1(c), measured by an optical spectrum analyzer (YOKOGAWA, AQ6317B) with a resolution of 0.01nm, where a narrow laser line sits on a broad amplified spontaneous emission pedestal. The linewidth is measured to be 0.128 nm. The ARHCF has a single-ring node-less structure consisting of 7 rings with 16.1 μm diameter and wall thickness of 323 nm, forming a negative-curvature hollow-core region with a diameter of 32.8 μm, as shown in the inset of Fig. 1(d) [29]. The attenuation spectrum of the ARHCF is simulated based on the finite-element method using the COMSOL software. Figure 1(d) shows the simulation result from 900 nm to 2500 nm wavelength. It can be seen that the ARHCF has a continuous transmission window with loss < 1dB/m up to 2 μm while an increased loss at longer wavelength. This allows the efficient generation of the 1st order vibrational Raman Stokes at 1.4 μm while blocking the generation of the 2nd and higher order vibrational Raman Stokes pulses above 2 μm. The ARHCF is sealed by two high-pressure custom gas cells, and coiled with a diameter of ~40 cm. Coated optical windows are assembled on both sides of the gas cells with >99% transmission at both 1.06 μm and 1.4 μm. $N_2$ is filled into the ARHCF through one of the two gas cells. A 1240 nm long-pass filter (FELH1240, Thorlabs) is placed at the output to separate the Raman laser from the residual pump.

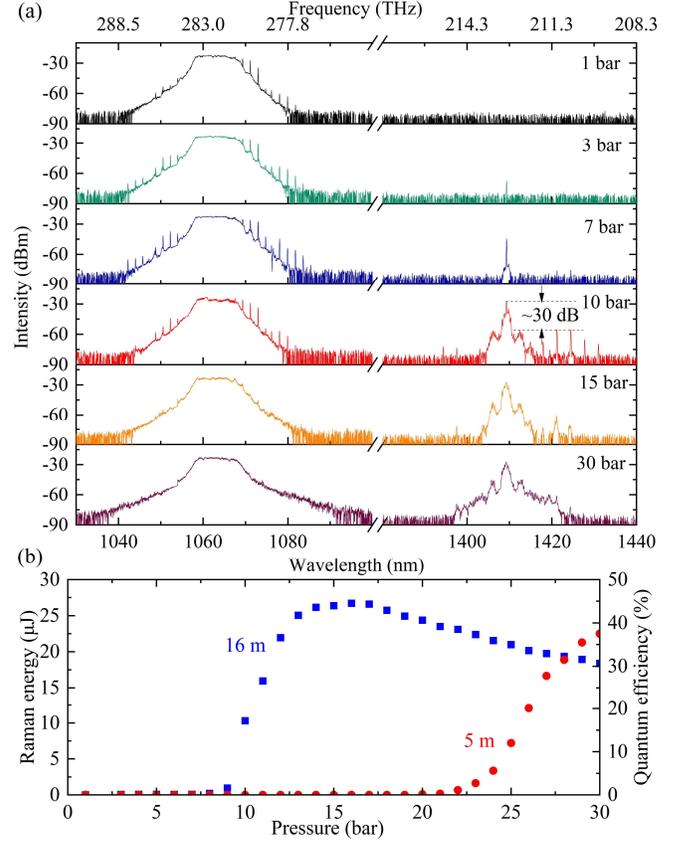

Fig. 2. (a) Spectrum (logarithmic scale) evolution versus $N_2$ pressure. (b) Evolution of pulse energy and quantum efficiency of the vibrational Raman laser at 1.4 μm against pressure using two different lengths of ARHCFs.

The spectral evolution as a function of $N_2$ pressure is investigated at the maximum pump pulse energy and proper orientation of the HWP, as shown in Fig. 2(a). Each spectrum was measured 5 minutes after adjusting the $N_2$ pressure, to avoid pressure gradient within the ARHCF. At atmospheric (1 bar) $N_2$ pressure, a series of rotational Raman (anti-)Stokes lines with ~1.7 nm spacing are observed at the pump region. The pump spectral region shows a flat top, because in this measurement a band pass filter is placed before the input side of the ARHCF, to suppress ASE and to visualize the rotational (anti-)Stokes lines. When the $N_2$ pressure increases to 3 bar, the 1st order vibrational Raman line at 1.409 μm wavelength appears in the measured spectrum. Then, the intensity of the 1st order vibrational Raman Stokes increases by increasing the $N_2$ pressure. When the vibrational Raman Stokes line becomes sufficiently intense, it leads to the generation of several weak rotational Raman (anti-)Stokes lines with ~3.3 nm spacing around 1.4 μm region. The intensity ratio between the vibrational Raman line and its surrounding rotational Raman lines is measured to be >~30 dB. With the increase of the vibrational Raman line intensity, the rotational (anti-)Stokes lines around the pump line become weak because the gain coefficient of the vibrational SRS increases toward high $N_2$ pressure and depletes the main pump energy. The 2nd order vibrational Raman line at 2.1 μm was not observed due to the high ARHCF loss.

Figure 2(b) shows the measured vibrational Raman energies at different N₂ pressures. The energies of the rotational Raman lines around the vibrational Raman line is neglected due to its weak intensity. Because the Raman gain coefficient is dependent on the N₂ pressure, initially the Raman pulse energy increases with the increase of N₂ pressure and reaches a maximum value of 26.5 μJ at 15 bar, corresponding to a quantum efficiency of 45 %, which takes into account the 20% coupling loss from the pump laser to the ARHCF. At higher N₂ pressure, the pulse energy drops slowly, because the vibrational Raman Stokes reaches its maximum pulse energy at a shorter fiber length and then it experiences more fiber loss than vibrational Raman Stokes gain in the remaining length of the ARHCF. This process is validated with the energy evolution of the Raman pulse using an ARHCF with the same parameters but 5 m long length instead, as shown with the red dots in Fig. 2(b). In this case, since the Raman gain is always higher than the fiber loss, the pulse energy keeps increasing as the pressure increases, reaching 22 μJ at 30 bar, which is the maximum pressure of our setup.

two peaks of 20.47 kW and 9.19 kW separated by a gap of 2.7 ns. The durations of these two peaks are measured to be 260 ps and 390 ps, respectively, indicating the short dephasing time of the vibrational SRS effect in N₂. The peak at the leading edge has a longer duration and higher intensity than the peak at the trailing edge. This is linked to the transition process from the spontaneous Raman scattering to the SRS regime [30,31]. The inset of Fig. 3(b) shows the measured beam profile of the vibrational Raman laser using a beam profiler (BP109-IR2, Thorlabs). The Gaussian-like distribution indicates that it operates at the fundamental mode of the ARHCF. It is worth noting that the wavelength of this vibrational Raman laser line can be precisely tuned from 1407.5 nm to 1409.5 nm (~2 nm range) by thermally tuning the laser diode seed of the pump laser from 1060 nm to 1061 nm in the temperature range of 20-35 °C, which is necessary for high-resolution gas sensing application. Figure 3(c) shows the measured center wavelengths of both the pump and the corresponding vibrational Raman Stokes as a function of temperature. Both of them show a linear dependence on the temperature of the laser diode seed.

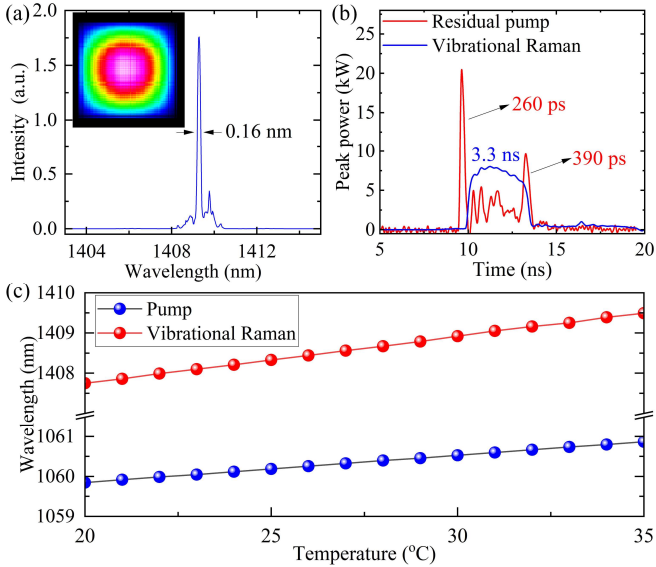

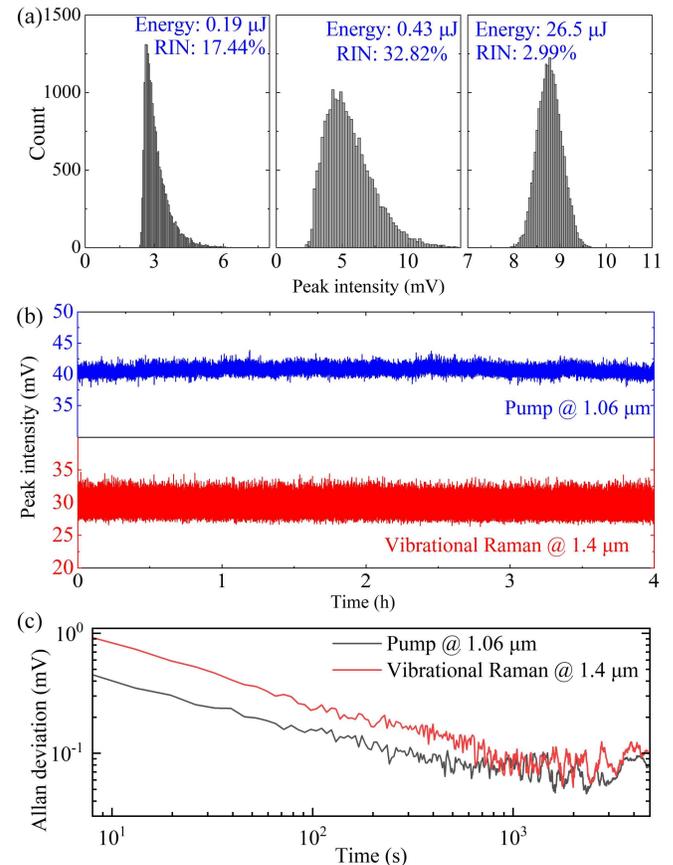

Fig. 3 (a) Measured spectrum (linear scale) of the vibrational Raman Stokes line at 10 bar N₂ pressure. The inset shows the measured beam profile of the vibrational Raman laser. (b) Pulse profiles of the residual pump and the vibrational Raman laser measured at 15 bar N₂ pressure. (c) Wavelength tunability of the pump and the corresponding vibrational Raman laser as a function of laser diode seed temperature.

The FWHM linewidth of the vibrational Raman was measured to be 0.16 nm, as shown in Fig. 3(a), which is slightly dependent on N₂ pressure. Figure 3(b) shows the measured typical pulse profile of both the residual pump and the vibrational Raman pulse at 15 bar N₂ pressure. The residual pump is measured by extracting it from the vibrational Raman beam with a 1300 nm short-pass filter. Both pulse profiles are normalized by their measured average pulse energies (26.5 μJ for the vibrational Raman and 4.8 μJ for the residual pump), to show their peak powers. As a result, it can be seen that the vibrational Raman Stokes exhibits a peak power of 8.03 kW with a pulse duration of 3.3 ns. The residual pump exhibits

Fig. 4. (a) Measured histograms of peak intensities of the pump and the vibrational Raman laser. (b) Long-term stability monitoring of the pulse peak intensity of the pump (top) and the vibrational Raman lasers (bottom). (c) Allan deviation calculated from (b). N₂ pressure in this measurement is 15 bar.

We also investigated the noise performance and long-term stability of the vibrational Raman laser at 1.4 μm. The noise performance is evaluated in terms of relative intensity noise (RIN) of the pulse peak intensity measured with the 5 GHz near-IR photodetector (DET08C/M, Thorlabs). Figure 4 (a) shows the measured histograms of the peak intensity distribution at three different Raman pulse energies of 0.19, 0.43, and 26.5 μJ, respectively. The distribution shows a negative exponential shape with a high RIN of 17.4 % at the low pulse energy case of 0.19 μJ, which is a typical characteristic of SRS where the Raman laser is initiated from quantum noise [32]. When the energy increases, the distribution transits gradually toward a Gaussian-like shape with a lowered RIN, which is related to the significant depletion of the pump energy. The RIN is measured to be 2.99% at the maximum vibrational Raman pulse energy of 26.5 μJ, which is higher than the pump RIN of 0.8 %. Figure 4 (b) shows the measured long-term stability of both the pump and the vibrational Raman laser over 4 hours in terms of the pulse peak intensity. This measurement was conducted after two hours of warming-up of the laser. It can be seen that the peak intensity of the pump and the Raman pulses remains stable over time. To better describe the long-term stability, Fig. 4(c) shows the calculated Allan deviation of the data in Fig. 4(b). It can be seen that, in the logarithmic scale, the Allan deviation of both the pump and the Raman laser show a linear decreasing trend over a long time period of ~1000 s, indicating that the laser has good long-term stability (i.e., without drift) [33]. This property is important for trace-gas sensing, because the white-noise fluctuation of the pulse peak intensity can be well mitigated through averaging algorithm [34]. This good long-term stability is attributed to the low heat release due to the low pump average power and the low quantum defect during the vibrational SRS wavelength conversion from 1 μm to 1.4 μm [32].

In conclusion, a high energy and narrow linewidth Raman laser with emission wavelength at 1.4 μm is generated based on the vibrational SRS effect in a 16 m long $N_2$-filled ARHCF. The maximum Raman pulse energy of 26.5 μJ is obtained with an $N_2$ pressure of only 15 bar. The wavelength of the vibrational Raman laser is precisely tunable over 2 nm by thermally tuning the wavelength of the seed on the Yb-doped fiber laser pump around 1060 nm. Moreover, this work reveals the potential of further extending the wavelength toward longer near-infrared and mid-infrared regions, which is promising for high-resolution gas sensing [29]. For instance, the developed Raman laser combing high pulse energy with narrow linewidth could be used as a pump for another stage gas-filled ARHCF for further Raman wavelength conversion. Another potential solution is to select other well-established pump wavelengths (e.g., Er-doped fiber laser at 1.55 μm, and Tm-doped fiber laser at 2 μm) for $N_2$-filled ARHCF Raman laser generation. The proposed laser also holds the potential to be developed into a compact fiber structure by splicing the ARHCF with the pump fiber laser [35–37].

**Funding.** This work is supported by VILLUM Fonden (Grant No. 36063, Grant No. 40964), LUNDBECK Fonden (Grant No. R346-2020-1924), and US ARO (Grant No. W911NF-19-1-0426).

**Acknowledgments.** We thank Martin Nielsen (affiliated with DTU Space) for fabricating the gas cells. Lujun Hong now is affiliated with: Institute of Space Science and Technology, Nanchang University, Nanchang 330031, China.

**Disclosures.** The authors declare no conflicts of interest.

**Data availability.** Data underlying the results presented in this paper are available upon request.